\documentclass[acmsmall]{acmart}
\usepackage{fontenc}
\usepackage{wrapfig}

\newcounter{RQCounter}
\newcounter{HCounter}
\newcounter{RSCounter}
\newcommand{\RQ}[2]{%
\refstepcounter{RQCounter} \label{#1}
 	\vspace{0.1in}
    \noindent \textbf{RQ\arabic{RQCounter}.~#2
	\vspace{0.05in}
    }
}

\usepackage{framed}

\newcommand{\numRepos}{18,247\xspace}
\newcommand{\numReadmes}{350,308\xspace}
\newcommand{\numRoundOneReadmes}{24,767\xspace}

\usepackage[normalem]{ulem}

\newcommand{\mysec}[1]{\vspace{0.1cm} \noindent \textbf{\textit{#1:}}}

\usepackage{setspace}

\usepackage{booktabs}
\usepackage{array}
\newcommand{\PreserveBackslash}[1]{\let\temp=\\#1\let\\=\temp}
\newcolumntype{C}[1]{>{\PreserveBackslash\centering}p{#1}}
\newcolumntype{R}[1]{>{\PreserveBackslash\raggedleft}p{#1}}
\newcolumntype{L}[1]{>{\PreserveBackslash\raggedright}p{#1}}
\newcolumntype{Y}[1]{>{\linespread{.5}}p{#1}}

\usepackage{xcolor}
\definecolor{brickRed}{HTML}{D1495B}

\newcommand{\bv}[1]{{\color{brickRed}\bfseries BV: #1}}
\usepackage{tcolorbox}

\usepackage{caption}
\usepackage{subcaption}
\usepackage{tabularx}
\usepackage{booktabs} 
\usepackage{multirow}

\usepackage{xspace}

\usepackage{graphicx}
\usepackage{todonotes}
\usepackage{url}
\usepackage{soul}
\usepackage{booktabs}
\usepackage{longtable}
\usepackage{supertabular}
\usepackage{multicol}
\usepackage{lipsum}     


\usepackage{hyperref}

\begin{document}

\widowpenalty=10000
\clubpenalty=10000

\title[Scientific Open-Source Software Is Less Likely to Become Abandoned Than One Might Think...]{Scientific Open-Source Software Is Less Likely to Become Abandoned Than One Might Think! Lessons from Curating a Catalog of Maintained Scientific Software}

\author[A. Malviya Thakur]{Addi Malviya Thakur}
\orcid{0000-0002-2681-9992}
\affiliation{%
  \institution{University of Tennessee at Knoxville}
  \city{Knoxville}
  \country{USA}
}
\affiliation{%
  \institution{Oak Ridge National Lab}
  \city{Oak Ridge}
  \country{USA}
}
\author[R. Milewicz]{Reed Milewicz}
\orcid{0000-0002-1701-0008}
\affiliation{%
  \institution{Sandia National Laboratories}
  \city{Albuquerque}
  \country{USA}
}
\email{rmilewi@sandia.gov}

\author[M. Jahanshahi]{Mahmoud Jahanshahi}
\orcid{0000-0003-4408-1183}
\affiliation{%
  \institution{University of Tennessee at Knoxville}
  \city{Knoxville}
  \country{USA}
}
\email{mjahansh@vols.utk.edu}

\author[L. Paganini]{Lavínia Paganini}
\orcid{0000-0002-2729-0314}
\affiliation{%
  \institution{Eindhoven University of Technology}
  \city{Eindhoven}
  \country{Netherlands}
}
\email{l.f.paganini@tue.nl}

\author[B. Vasilescu]{Bogdan Vasilescu}
\orcid{0000-0003-4418-5783}
\affiliation{%
  \institution{Carnegie Mellon University}
  \city{Pittsburgh}
  \country{USA}
}
\email{vasilescu@cmu.edu}

\author[A. Mockus]{Audris Mockus}
\orcid{0000-0002-7987-7598}
\affiliation{%
  \institution{University of Tennessee at Knoxville}
  \city{Knoxville}
  \country{USA}
}
\email{audris@utk.edu}









\begin{abstract}

  Scientific software is essential to
  scientific innovation and in many ways it is distinct from other
  types of software. Abandoned (or unmaintained), buggy, and hard to use
  software, a perception often associated with scientific software
  can hinder scientific progress, yet, in contrast to other types of
  software, its longevity is poorly understood.  
  Existing data curation efforts
  are fragmented by science domain and/or are small in scale and
  lack key attributes.  We use large language models to classify 
  public software repositories in World of Code into distinct 
  scientific domains and layers of the software stack, curating a large and
  diverse collection of over 18,000 scientific software projects.
  Using this data, we estimate survival models to understand how
  the domain,
  infrastructural layer, and other attributes of scientific
  software affect its longevity. We further obtain a matched sample
  of non-scientific software repositories and investigate the
  differences.  We find that infrastructural layers, downstream
  dependencies, mentions of publications, and participants from
  government are associated with a longer lifespan, while newer
  projects with participants from academia had shorter
  lifespan. Against common expectations, scientific projects have a
  longer lifetime than matched non-scientific open-source software
  projects. We expect
our curated attribute-rich collection to support future research on scientific software and provide
insights that may help extend longevity of both scientific and other projects.

\end{abstract}

\maketitle

\section{Introduction}
\label{sec:introduction2}
\looseness=-1
Computing is central to science,
enabling next-generation simulations and data analyses that drive innovations in fields like medicine, energy, climate science, and engineering~\cite{cerf2024boundary}. 
These advances
all depend
on an ever-growing ecosystem of open-source and community-driven scientific software. In the past two decades, the scientific software community has largely moved away from privately-developed software used by small teams, and towards open science and open data as well as FAIRness (findability, accessibility, interoperability, and reusability)~\cite{ramachandran2021open, lamprecht2020towards, fouilloux2023building}.



As this ecosystem has grown, however, so too have concerns about its sustainability, understood broadly 
as its \textit{ability to endure}~\cite{becker2015sustainability}, or to be maintained in a state where it continues to provide value to the scientists using it (e.g., \cite{trainer2014community, morris2021understanding}).
There are many challenges to sustaining open-source software (OSS)~\cite{chengalur2010sustainability, valiev2018ecosystem, coelho2017modern}, to which scientific OSS adds a unique set.
These include insufficient training in software development on the part of researcher-developers~\cite{howison2011scientific}, interdisciplinary collaboration challenges~\cite{sun2024sustain}, mismatched priorities of science funders that tend not to value software work highly~\cite{johanson2018software}, and a need for highly specialized skills for ongoing maintenance to not only maintain compatibility, portability, etc., but also scientific relevance~\cite{rechert2021long}. In short, progress has been promising but scaling sustainable and trustworthy OSS software still hinders scientific advances. 


However, while there is increasing recognition of the importance of scientific OSS and of the challenges to sustaining this digital infrastructure, we lack a centralized view of this infrastructure and its state of maintenance. We also lack empirical evidence on the key factors that influence the health and sustainability of scientific OSS. 
Both are critically needed, e.g., to draw attention to critical scientific OSS projects that are undermaintained and at risk of becoming abandoned, and to help guide resource allocation.

\begin{figure}
\vspace{-5mm}
    \centering
    \includegraphics[width=0.90\linewidth, clip=true, trim = 0 40 0 0]{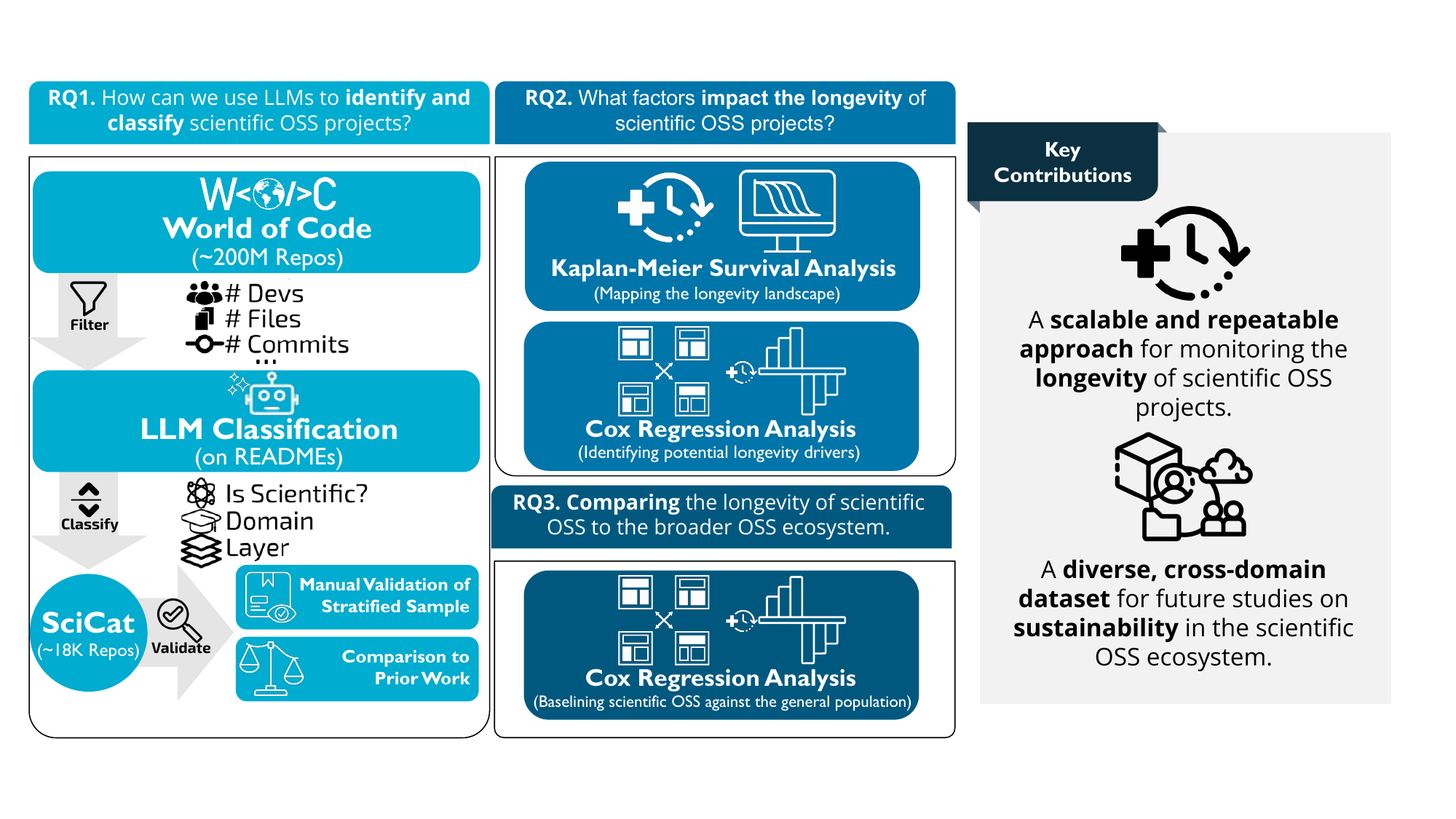}
    \caption{A visual walkthrough of the methodology used in our paper.}
    \label{fig:methodologyWalkthrough}
\vspace{-2mm}
\end{figure}

To this end, in this paper, we report on a two-part empirical study (Figure~\ref{fig:methodologyWalkthrough}). 
First, we present a novel approach to identify scientific OSS repositories and classify them in terms of scientific domain and software stack layer (from small-scale applications to infrastructure), resulting in a dataset of \numRepos repositories. 
Second, we conduct survival analyses of these 
repositories, modeling the factors associated with a higher risk of becoming abandoned and estimating how the risk of becoming abandoned compares to a matched set of non-scientific OSS projects. 

Our analysis reveals several key insights:
Many scientific software repositories lack mentions of publications or funding, 
and thus risk being overlooked. 
Mathematics-related projects exhibit the longest average lifespan, whereas projects in computer science have the shortest. 
Scientific infrastructure-layer projects demonstrate the greatest longevity, supported by community involvement, more downstream users, and government participation. Projects with more upstream dependencies and academic involvement tend to have shorter lifespans. And finally, scientific OSS tends to survive longer than non-scientific counterparts.

In short, the contributions of our work are: (1)~a validated
AI-assisted methodology for detecting scientific software
repositories at scale, (2)~a cross-cutting dataset of \numRepos
scientific software repositories spanning different scientific
domains and technology stack layers, (3)~quantitative findings on
scientific software longevity including survival curves and
proportional hazards models, and (4)~a comparison between scientific OSS and generic OSS repositories.


\section{Background}
\label{sec:background}
Before we present our research questions, we start with some background on the scientific software ecosystem and the challenges to its sustainability, to set the stage.

\vspace{-0.3cm}
\subsection{The Scientific Software Ecosystem}

\looseness=-1
\textbf{Scientific software} broadly refers to software that is used for scientific purposes (see \cite{KANEWALA20141219}), e.g.,
domain-specific simulations, 
numerical libraries, visualization tools, and workflow managers~\cite{sochat2022research}.
It plays a central role in the scientific enterprise and drives discoveries that advance our well-being, prosperity, security, and understanding of the universe.

Scientific software development has had a long and storied history~\cite{dyson2012turing},
but largely independent from conventional software development \cite{faulk2009scientific}. Early-days scientific software was, as a rule, developed in private by small independent teams of domain scientists and mathematicians without formal training in software engineering \cite{johanson2018software}. 
In the past decade and a half, a convergence of different factors has forced the scientific community to revisit how it develops software; these include a crisis of lost productivity and credibility of results due to poor software quality~\cite{johanson2018software}, scientific challenges that cannot be solved without integrated multidisciplinary solutions~\cite{ober2017patterns}, and a shift toward greater openness in science through the sharing of code and data~\cite{ramachandran2021open}. 
This has led to scientific software becoming increasingly (1)~community-driven (\textit{c.f.,} \cite{bonomi2019promoting, charles2020toward}), (2)~open source~\cite{katz2018joss, hasselbring2020open}, and (3)~produced both by researchers who have a growing interest in software engineering best practices~\cite{adorf2018professionally, lee2018ten, haider2021realization, queiroz2017good, dubey2020distillation} and a growing number of research software engineering (RSE) professionals~\cite{baxter2012research, brett2017research, cohen2020four}. 

Today, scientific software is part of a predominately open-source \textbf{software ecosystem} (or SECO) --- also referred to by \cite{jansen2022fairseco} as the Worldwide Research Software Ecosystem (WRSE). 
Here, we use the definition by \cite{manikas2013software}, which is that an ecosystem encompasses ``the interaction of a set of actors on top of a common technological platform that results in a number of software solutions''. 
That is, we have a diverse range of \textbf{actors} (individual scientific software developers, teams, and institutions) coordinating their work on common \textbf{platforms} (\textit{e.g.,} GitHub, Gitlab, and Bitbucket) to develop software \textbf{solutions} for scientific applications (\textit{e.g.,} multi-physics simulation codes, high-performance computing libraries, and visualization tools) which are assembled into complex software stacks by users (who are frequently developers themselves). 

According to Hinsen's conceptual model of scientific software stacks~\cite{Hinsen2019}, 
scientific software developers may contribute packages at different \textbf{layers}: 
the topmost layer consists of \textbf{project-specific code}, more likely written by individual scientists, and in an ad-hoc manner for a particular research project; one layer below are \textbf{domain-specific tools}, which can vary in their level of maturity and user base; finally, below there is \textbf{infrastructure} created specifically for scientific computing, which is expected to be mature and to serve multiple fields of research.

\vspace{-0.2cm}
\subsection{(Un)sustainability in the Scientific Software Ecosystem}
\label{subsec:sustainabilityBackground}

As the scientific software community has matured, the shift towards a complex, interdependent, and open-source SECO has also introduced new challenges, especially concerning software \textbf{sustainability}. 
By sustainability, we follow the the Karlskrona Manifesto to mean the ability of a socio-technical system (such as a software project) to endure, a concept which spans individual, social, technical, economic, and environmental dimensions~\cite{becker2015sustainability}. 
However, we focus on the \textbf{technical} and \textbf{economic} dimensions of sustainability,\footnote{This is not to suggest that environmental, individual, and social dimensions of sustainability are of lesser concern.}
as these align most directly with another common view of sustainability in the software literature, i.e., ``the ability to maintain the software in a state where scientists can understand, replicate, and extend previously reported results that depend on that software''~\cite{trainer2014community}. 
In this sense, the \textbf{longevity} of information, systems, and infrastructure is a prerequisite for their adequate evolution with changing surrounding conditions (technical) and maintaining capital and added value (economic), respectively.

As discussed above, the scientific software community has moved towards closer coordination and greater reuse of each other's open-source software. That shift exposes projects to the usual OSS sustainability problems plus science-specific pressures. 
In conventional OSS SECOs, packages can become obsolete or inactive~\cite{cogo2021deprecation, miller2025npm}, breaking changes can cause cascading effects on projects~\cite{bogart2021breaking}, communities around software projects can rise and fall~\cite{valiev2018ecosystem}, projects can die due to lack of financial viability~\cite{xavier2020software}, and poor governance can cause an ecosystem to splinter and fall apart~\cite{gamalielsson2014sustainability}; the scientific software ecosystem is much the same in these respects. 
However, unlike the creation or maintenance of OSS in and of itself, many scientific applications have to be regularly ported to new and cutting-edge hardware (necessitating costly rework)~\cite{rechert2021long}, when packages are revealed to be untrustworthy and irreproducible it can jeopardize any research based on them (possibly harming the careers of the users!)~\cite{soergel2014rampant}, and much of the scientific software in existence was created by people whose job is to do science, \textit{not} to write software~\cite{johanson2018software}. 
These dynamics concern the public and funders who support and nurture this ecosystem~\cite{barker2022overview, strasser2022ten}; without a holistic perspective, they run the risk of funding duplicative projects, failing to grow communities around software to carry it forward, or propping up poor quality software that collapses under its own technical debt. At present, however, there is relatively little data on scientific software sustainability at an ecosystem level, which impairs effective decision-making.

In recent years there have been a range of works on scientific software sustainability, including experience reports and case studies~\cite{sun2024sustain}, reports on developers' practices and perceptions of sustainability~\cite{feitosa2023understanding}, and community-building strategies~\cite{ram2018community}. On the repository mining front, studies have investigated different proxies for sustainability, such as cross-project references~\cite{sun2024sustain} and complexity metrics~\cite{willenbring2021evaluating}. These studies, however, are relatively small scale and domain specific.

\looseness=-1
In contrast, we provide a baseline of data to support understanding and awareness of the broader scientific OSS ecosystem. 
For that reason we focus our attention on \textbf{longevity} of projects and correlates (\textit{e.g.,} community size, number of dependencies, etc.), as these are well-attested in the SECO literature and allow us to compare scientific OSS to other kinds of OSS documented in prior work. 
Given that the transferability of findings from the conventional OSS SECO literature to the scientific software domain is uncertain, having this data allows us to make well-grounded comparisons.

While this study focuses on STEM disciplines, scientific software is also essential in domain such as social sciences, arts, and humanities. For example, fields such as linguistics, archaeology, history, and political science rely on computational tools for text analysis, geospatial mapping, and data mining~\cite{arnold2024humanities, 10.1093/oso/9780197756874.001.0001}. However, these disciplines often have different funding models, collaboration structures, and sustainability challenges~\cite{Newfield_2025}.





\section{Research Questions}
\label{subsec:rqs}

Our study consists of two parts. First, we take a representative cross-section of the scientific OSS ecosystem, which requires a new technique to identify and categorize the software projects, as no such dataset exists. Second, we study the factors associated with increased longevity (or, conversely, increased risk of abandonment) within our sample and across scientific and non-scientific OSS.
Next we formulate and motivate each of our research questions.

\RQ{rq1}{How can we use LLMs to identify and classify scientific OSS projects?}

\noindent \textbf{\textit{What We Know.}} Ours is not the first study to assemble a corpus of scientific software projects. Several efforts have been made to build small datasets of projects to study particular phenomena such as security~\cite{murphy2020curated}, contributor roles~\cite{milewicz2019characterizing}, sustainability-related practices and incentives~\cite{Howison2011}, and tools for metadata extraction~\cite{mao2019somef}. Other studies have sought to build larger datasets by using publication metadata (e.g., citations; and metadata entries in publications, grant-funding-related archives, and institutional catalogs)~\cite{10.1145/3643991.3644876, kelley2021framework, wattanakriengkrai2022github}. 
The closest work to ours in terms of scale and methodology is by \citet{wattanakriengkrai2022github}, who scanned GitHub for repositories with links to academic papers, turning up 20,278 links. 
Finally, there have been several efforts in recent years to make scientific software more discoverable such as \href{https://paperswithcode.com/}{Papers with Code} which catalog and promote code repositories associated with publications.

\smallskip
\noindent \textbf{\textit{What We Do Not Know.}} Prior works, while all valuable, (1)~have focused on populations that are very small, (2)~target a specific domain, (3)~do not break down projects according to layers, 
or (4)~rely on some kind of self-reporting such as registering with an online catalog or GitHub links in publications. We hypothesize, however, that both publication and citation practices as well as software development practices may differ across scientific domains and layers of the software stack. This led us down a different path, which is to detect scientific software repositories online based on their content. In particular, we investigate whether large language models (LLMs) could assist with detecting and classifying these repositories.

\RQ{rq2}{What factors impact the longevity of scientific OSS projects?}

\noindent \textbf{\textit{What We Know.}} 
It is argued that ``the lifespan of scientific software tends to be either very long or very short''~\cite{sanders2008development}. Exact numbers, however, are harder to come by; as \citet{Hinsen2019} has noted, the ``uncertain survival probability'' of any given project complicates the decision-making for all actors in the ecosystem. The works most closely related to ours are a pilot study by \citet{hasselbring2020open} and follow-up work by \citet{eitzen2020research}, who measured longevity of scientific software projects found by searching for mentions of DOIs/citations in repositories and by URLs in papers. Among their findings, \cite{hasselbring2020open} report significant differences between domains in terms of longevity; in particular, they report that the median lifespan for computational science software repositories in their dataset is 15 days vs.\ 5 years for computer science repositories.

\smallskip
\noindent \textbf{\textit{What We Do Not Know.}} 
As both \citet{hasselbring2020open} and \citet{eitzen2020research} include one-off repositories for scripts associated with papers in their samples, it is unclear how previous findings would generalize to scientific software projects that are intended to be actively developed and maintained, and that may have a large dependent user base.
While we recognize that there are sustainability concerns involved in those one-off paper repositories (e.g., for reproducibility of scientific results), arguably becoming under- or unmaintained does not pose as much of a threat to the stability of the ecosystem as would, say, a core math library shutting down.
Thus, we focus on scientific OSS projects that are more likely intended to endure, and ask the following subquestions:

\begin{description}
    \item \textbf{RQ2a.} How long do scientific OSS projects survive, and how does this vary across different domains and layers of the software stack?
    \item \textbf{RQ2b.}
    What factors predict the longevity of scientific OSS projects?
\end{description}

Longevity is a prerequisite for the much more complex notion of sustainability. 
With occasional exceptions of ``feature-complete'' projects~\cite{valiev2018ecosystem}, software systems need continuous maintenance to remain compatible with changing environments, fix bugs and security vulnerabilities, add needed features, etc.
In OSS, such maintenance cannot be taken for granted, as contributors and maintainers are often volunteers, and typically free to disengage at any time.
All these maintenance updates, at the very least, require a project to be active (``alive''), as opposed to dormant or abandoned.\footnote{Although being ``alive'' is not sufficient for a project to be well-maintained, as the available maintenance effort may still fall short of the demand for updates~\cite{champion2021underproduction}.}
Thus, longevity is a critical indicator of health and success in both traditional~\cite{zhou2016inflow, valiev2018ecosystem} and scientific open-source software~\cite{jorgensen2021}. 

\RQ{rq3}{How does the longevity of scientific OSS compare to the broader OSS ecosystem?}

\noindent \textbf{\textit{What We Know.}} As discussed in Section~\ref{sec:background}, there are (1)~clear differences that distinguish scientific software development from its conventional counterpart, but also that (2)~as the scientific software ecosystem has grown, it has had to contend with problems shared by many other OSS ecosystems. 

\smallskip
\noindent \textbf{\textit{What We Do Not Know.}} 
The challenges that are unique to scientific OSS may change the dynamics of the scientific software ecosystem in unpredictable ways, possibly threatening the sustainability of scientific OSS even further. Given the importance of scientific open-source software for science, it is important to better understand how it compares to other types of OSS to have a better chance of identifying undermaintained projects in need of additional support.
To our knowledge, there are no studies to date that have empirically compared the OSS scientific software ecosystem to any non-scientific counterparts.
From a software engineering research perspective, being able to compare and contrast scientific and non-scientific software at an ecosystem level would be very helpful, as it builds on the wealth of OSS ecosystem research that is already available. 
\section{RQ1: Identifying and Classifying Scientific OSS Software} 
\label{sec:RQ1}
We start by presenting our novel approach to identify scientific OSS repositories based on their contents and classify them in terms of domain and Hinsen's software stack layers~\cite{Hinsen2019}.
\vspace{-1mm}

\subsection{Methods}
The main goal of our approach is to build a large, multi-domain, multi-layer dataset of scientific software with public code repositories that we will use to study scientific software longevity in different contexts at scale. 
To this end, we developed and validated an automatable, content-centric approach to identify and categorize scientific software among OSS repositories indexed by the World of Code (WoC) research infrastructure~\cite{ma2021world}. Our method uses a large language model (LLM) prompted with the contents of the README files of the repositories. READMEs typically contain information on what a project does and how it works~\cite{prana2019categorizing}, and we expect this information can identify scientific software even when no explicit links to publications or acknowledgments of grants are present, which is an identification approach used in prior work.



At a high level, our process consists of two parts: (a)~sampling OSS repositories using the WoC infrastructure and (b)~using an LLM to categorize them.

\mysec{Sampling and Pre-processing}
WoC contains metadata for nearly all public software projects, including GitHub, GitLab, and BitBucket, as well as numerous smaller platforms and individual forges. 
Among others, the WoC data are curated to identify and cross-link forks~\cite{forks20} and different aliases corresponding to unique author IDs~\cite{identity20}, and include information on all versions of the code (including READMEs), commit activity timelines, and time-stamped package dependencies, which are needed for our analysis.
When we collected our data, WoC was at version \textit{V}, which was updated with new repositories found between March 1--30, 2023 and git objects retrieved by mid-May 2023; it included over
209 million repositories (including forks), ``deforked'' to 131 million unique projects. 

Sampling was necessary for two reasons. First, conceptually, it is well known that many public repositories are not meant for software development, but rather contain code dumps, student homework assignments, and other types of artifacts for which questions of longevity and sustainability are less relevant~\cite{kalliamvakou2016depth, munaiah2017curating}. Second, pragmatically, at the time of our study it was simply infeasible to process all repository READMEs in WoC using an LLM. 

Since there are no universally accepted criteria for identifying ``real'' software projects among those with public repositories, we applied a set of commonly used size and activity-based heuristics grounded in the literature~\cite{carruthers2022software}. 
For example, previous research found that project maturity correlates with repository size~\cite{liao2019prediction} (larger repositories are more likely to correspond to actual software development projects) and having some minimum amount of activity~\cite{he2024revealing, ait2022empirical} (code dumps tend to involve only a few commits over a short period); 
there is also evidence that scientific software is often developed in collaborative settings by multiple authors~\cite{koehler2020better}. 
In the same spirit, our sampling strategy considers the number of files in the repository (at least 10), the number of commits (more than 300), the number of commit authors (at least 3), the number of months with activity in the project (more than 6 consecutive active months), the last commit date (after November 2018, to avoid less relevant repositories that may have been abandoned long ago), and the programming language(s) identified in the repository (to remove repositories without identifiable code). 

Through this process, we filtered the 131 million deforked projects in WoC down to 430,469 projects that met our criteria. 
We then filtered projects with top-level file names that matched the string ``readme'' (case insensitive), which captures the modern README.md convention commonly used by GitHub, as well as many other older variations.
This resulted in our filtered dataset of \numReadmes README files (one per repository) used for LLM classification. 


Note that it is possible that some scientific OSS projects did not meet our sampling criteria, and thus did not get a chance to be considered. Similarly, it is possible that some repositories not intended for active development and long-term maintenance remain in our sample despite our relatively strict filters.
Our goals here are not to be all encompassing (i.e., identify all scientific OSS) nor perfectly precise (indeed, we expect both goals are futile), but rather to compile a sufficiently large and diverse sample for our subsequent analyses. 
Importantly, our sampling criteria should not affect the soundness of our subsequent analyses, since we use the same filtering criteria to identify the matching non-scientific OSS repositories we compare against.

\mysec{Identifying and Classifying The Scientific Software}
We used prompt engineering on OpenAI's LLMs to determine whether the
repositories in our sample can be considered scientific software,
to identify the likely layer they occupy in
Hinsen's~\citeyearpar{Hinsen2019} scientific software stack
(discussed above), and the most likely scientific domain they are
intended for. 
At the time of our data collection, GPT~4 was the top performing model available.
However, the number of README files for classification (\numReadmes) was still larger than we could feasibly process using GPT~4 (in terms of both speed and cost per query).
Therefore, we first performed a less precise but more efficient first pass through GPT~3.5 to further narrow down our sample of candidate READMEs, and then made a second pass over this smaller sample using GPT~4 for final classification with expectedly higher accuracy.

\begin{wrapfigure}{R}{0.31\textwidth}
    \vspace{-0.2cm}
  \begin{center}
    \includegraphics[width=0.3\textwidth, clip=true, trim= 0 0 0 0]{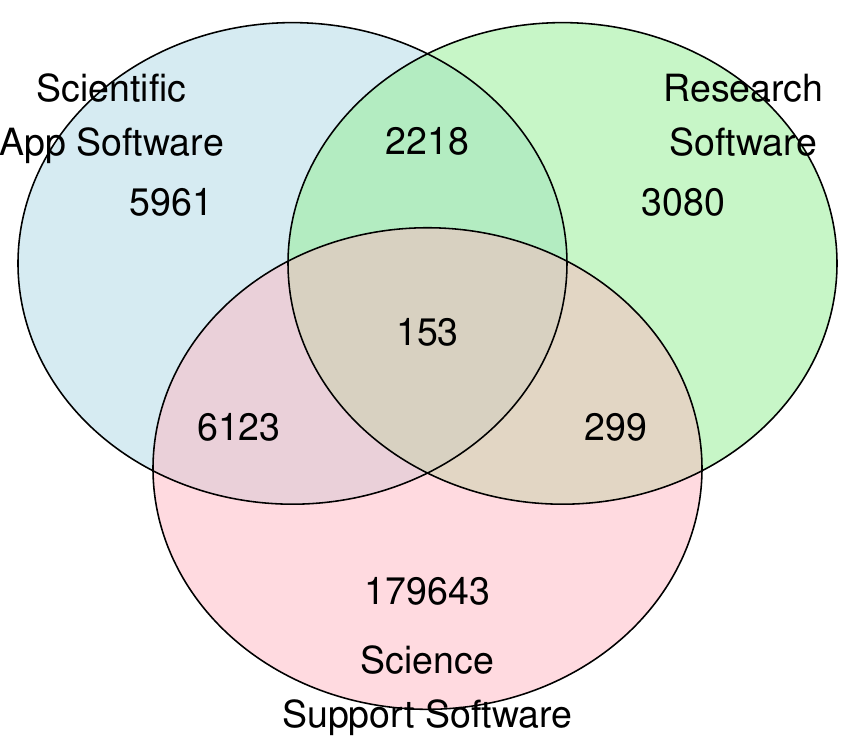}
  \end{center}
  \vspace{-0.2cm}
  \caption{\small Overlap between the three scientific-software-related labels assigned by GPT~3.5.\vspace{-0.2cm}}
  \label{fig:venn1}
\end{wrapfigure}
Concretely, for our first pass, we instructed GPT 3.5 to determine the repositories' relevance to scientific research based on the contents of their README. 
After five rounds of prompt engineering, we arrived at a prompt (see Supplementary Materials) asking the LLM to answer six questions about each README, including whether it describes \textit{scientific application software} (defined as ``domain-specific science and engineering software'' in our prompt), \textit{science support software} (i.e., ``software used to support scientific applications or research''), or \textit{research software} (i.e., ``software used to generate, process or analyze results that you intend to appear in a publication''); and whether it mentions a publication or research funding related to the software.
All three types describe software used for scientific purposes, i.e., can be considered \textbf{scientific software}~\cite{KANEWALA20141219}.
Including the latter questions about mentions of papers or funding in the prompt allows us to compare the content-based classification (the most novel component of our approach) with the traditional classification based on explicit links to publications or funding, which has been used in prior work (albeit not with LLMs).

\begin{wrapfigure}{R}{0.31\textwidth}
    \vspace{-0.7cm}
  \begin{center}
    \includegraphics[width=0.3\textwidth, clip=true, trim= 0 0 0 0]{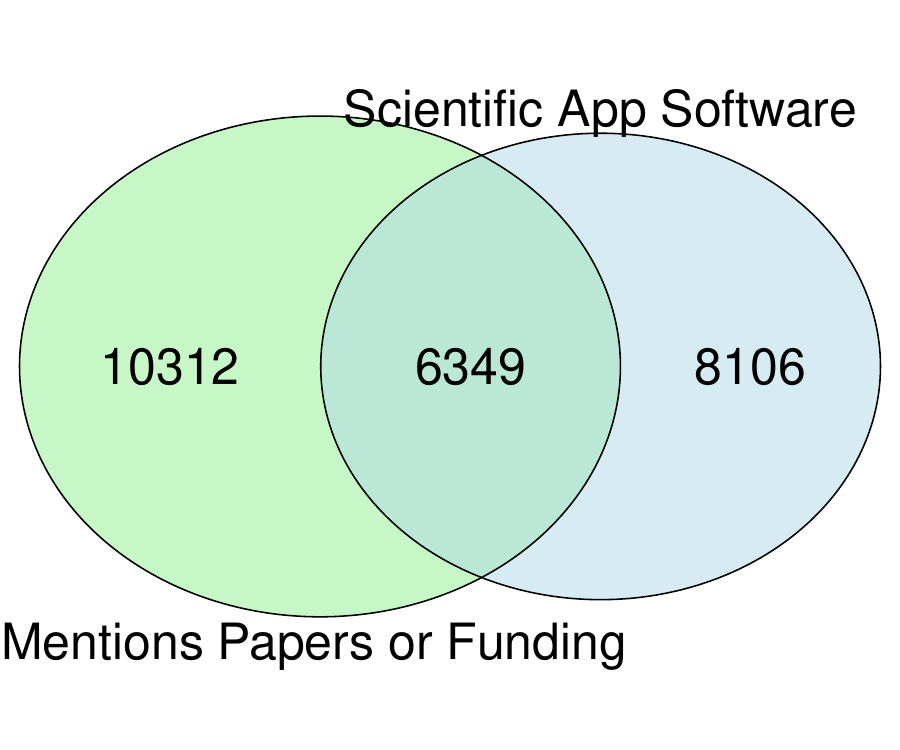}
  \end{center}
  \vspace{-0.2cm}
  \caption{\small Overlap between the scientific application software and paper/funding mention labels assigned by GPT~3.5.\vspace{-0.2cm}}
  \label{fig:venn2}
\end{wrapfigure}
As the definitions of the three types of software have high theoretical overlap, this step unsurprisingly yielded many READMEs tagged with multiple type labels (Figure~\ref{fig:venn1}), e.g., 43\% of the 14,455 repositories labeled as \textit{scientific application software} were also tagged as \textit{science support software}. 
Interestingly, while there is also considerable overlap between the \textit{scientific application software} group and READMEs labeled as mentioning a publication or research funding (Figure~\ref{fig:venn2}), our approach also identified many candidate scientific repositories without such explicit traces, that may not be discoverable using traditional approaches (we discuss the accuracy of our approach in the Evaluation section below).
Since the \textit{scientific application software} group had the highest overlap with the other two types and was of a more manageable size, we used these repositories for our second pass. In addition, we also included the remaining non-overlapping repositories with \textit{mentions of publications or research funding}, for a total of \numRoundOneReadmes repositories (READMEs) out of 342,656 successfully classified files.\footnote{7,654 prompts failed for various reasons, including server error, entity mismatch, rate limit failures, etc.}

In the second step, we prompted GPT 4 (see Supplementary Materials) to classify each of the \numRoundOneReadmes repositories into one of Hinsen's~\citeyearpar{Hinsen2019} seven \textbf{scientific software stack layers}, ranging from non-scientific infrastructure (layer 1) to non-research software (layer 7). 
We expected most software would fit into one of three science-related categories (i.e., layer 2: ``scientific infrastructure,'' layer 3: ``scientific domain-specific code,'' and layer 4: ``publication-specific code'') because of our first filtering pass.
However, we kept all seven layers in the prompt as we expected some level of inaccuracy from the first pass.
Additionally, as part of the same prompt, we asked the model to associate the repositories with one of 13 predefined \textbf{STEM fields}: Astronomy, Biology, Chemistry, Computer Science, Data Science, Earth Sciences, Engineering, Mathematics, Medicine, Neuroscience, Physics, Quantum Computing, and Statistics. 
While software is also vital beyond traditional scientific disciplines, e.g., in the social sciences and humanities, sustainability challenges in those contexts differ~\cite{tucker2022facing}, so we restrict the scope of our work to STEM fields to keep the analysis tractable. 
As above, we provided the contents of each README (truncated if needed to fit the LLM's context window) as part of the prompt.

\begin{wrapfigure}{R}{0.51\textwidth}
    \vspace{-0.4cm}
  \begin{center}
    \includegraphics[width=0.5\textwidth, clip=true, trim= 0 25 0 25]{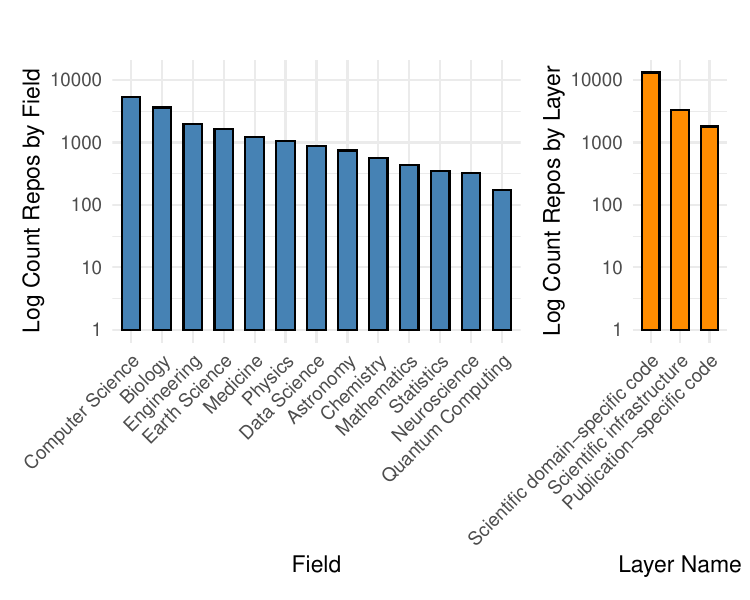}
  \end{center}
  \vspace{-0.2cm}
\caption{Distribution of the scientific repositories in our sample, by domain and scientific software stack layer.\vspace{-0.4cm}}
    \label{fig:distributions}
\end{wrapfigure}
Following this final labeling step, we discarded repositories not classified into one of the 13 STEM fields and three stack layers, arriving at our final \textbf{\textsc{SciCat}} dataset of \numRepos repositories (Figure~\ref{fig:distributions}). 
As can be observed from the figure, the dataset spans all stack layers and STEM fields, with repositories categorized as Computer Science being the most numerous. 
For example, these include security analysis tools for smart contracts like \href{https://github.com/ConsenSys/mythril}{\textsc{mythril}}, rendering tools like \href{https://github.com/tunabrain/tungsten}{\textsc{tungsten}} used by graphics researchers, 
and programming languages for high-performance computing like \href{https://github.com/taichi-dev/taichi}{\textsc{taichi}}; 
see dataset in replication package for more examples. 
Note, only one of the named examples mentions a publication in the README.

\vspace{-1mm}
\subsection{Evaluation}
To assess the accuracy of our classification, we used two strategies: (1)~manual validation by trained human raters on a sample of the data and (2)~cross-validation against existing datasets.

\mysec{Stratified Sample Validation}
We started with a pilot evaluation of the three LLM-generated labels (field, layer, and paper/funding mentions) for a stratified sample of 60 projects. 
Two human raters classified the same 60 projects on all three dimensions, and engaged in discussions to reconcile differences in their assessments. 
After resolving disagreements and re-labeling the same projects, the inter-rater reliability (IRR) for the categorization of STEM fields reached 0.733 and for software layers it reached 0.745, indicating ``substantial agreement''~\cite{landis1977application, fleiss1981measurement}. 
Most disagreements arose from projects in the machine learning domain, which can be applied across various fields. 
We also computed the IRR between human raters (after consensus) and the results generated by the LLM: these were 0.714 for field and 0.720 for layer, respectively.

Following the pilot, we expanded the manual validation process to a stratified sample of 468 projects in total, i.e., 12 repositories per field (across 13 fields) and per layer (across 3 layers), randomly sampled within each stratum. The three raters divided the remaining unlabeled projects between them and completed this categorization separately. In the end the IRR between human raters and the LLM was 0.650 for field and 0.624 for layer, both reflecting substantial agreement. 
For the classification of whether a project mentioned papers or funding, the IRR between the raters and the LLM was 0.467, indicating moderate agreement.

To further illustrate the agreement metrics, Table~\ref{tab:classification_metrics}, Table~\ref{tab:classification_metrics2} and Table~\ref{tab:mentionsPaperOrFunding} present the F1 score, precision, and recall for classification by field, layer, and mentions of papers or funding, respectively. 
For mentions of papers or funding, when the response is ``Yes'', the model achieved a high recall of 0.94 and an F1 score of 0.86, demonstrating strong performance in identifying true positives. However, for the ``No'' response, the recall and F1 decreased,
reflecting a tendency to miss instances where papers or funding were not mentioned.
Upon manual review of these instances, we found that the classifier tended to assign credit for the presence of any paper, whereas human raters evaluated whether the papers or citations were directly relevant to the code repository. When they were not, human raters correctly selected ``No'' as the answer. 
Overall, these findings highlight the reliability of the classification system but indicate the need for improved prompt engineering and contextual understanding in LLMs to better align with human judgment.

\begin{table}[t] 
\begin{minipage}{\textwidth}
\renewcommand{\arraystretch}{0.5} 
\setlength{\tabcolsep}{4pt} 
  \centering
  \footnotesize
  \setlength{\tabcolsep}{2pt}
    \begin{minipage}[t]{0.49\textwidth}
    \centering
    \captionof{table}{Classification metrics by field}
    \label{tab:classification_metrics}
    \begin{tabular}{lccc}
        \toprule
        Category & Precision & Recall & F1 Score \\
        \midrule
        Astronomy & 0.80 & 0.90 & 0.85 \\
        Biology & 0.79 & 0.79 & 0.79 \\
        Chemistry & 0.77 & 0.96 & 0.86 \\
        Computer Science & 0.86 & 0.40 & 0.55 \\
        Data Science & 0.50 & 0.53 & 0.52 \\
        Earth Science & 0.76 & 0.87 & 0.81 \\
        Engineering & 0.50 & 1.00 & 0.67 \\
        Mathematics & 0.86 & 0.81 & 0.83 \\
        Medicine & 0.68 & 0.82 & 0.74 \\
        Neuroscience & 0.78 & 0.88 & 0.82 \\
        Physics & 0.72 & 0.76 & 0.74 \\
        Quantum Computing & 0.83 & 0.97 & 0.90 \\
        Statistics & 0.76 & 0.81 & 0.79 \\
        \midrule
        Macro Avg & 0.64 & 0.70 & 0.66 \\
        Weighted Avg & 0.74 & 0.74 & 0.72 \\
        \bottomrule
    \end{tabular}
  \end{minipage}
\hfill
  \begin{minipage}[t]{0.49\textwidth}
    \centering
    \captionof{table}{Classification metrics by layer}
    \label{tab:classification_metrics2}
    \begin{tabular}{L{3.5cm}ccc}
        \toprule
        Category & Precision & Recall & F1 Score \\
        \midrule
        Publication-specific code & 0.81 & 0.92 & 0.86 \\
        Scientific domain-specific code & 0.83 & 0.62 & 0.71 \\
        Scientific infrastructure & 0.60 & 0.82 & 0.69 \\
        \midrule
        Macro Avg & 0.56 & 0.59 & 0.57 \\
        Weighted Avg & 0.76 & 0.75 & 0.74 \\
        \bottomrule
    \end{tabular}
    \vspace{0.2cm} 
    \captionof{table}{Classification metrics for paper/funding mentions} 
    \label{tab:mentionsPaperOrFunding}
    \begin{tabular}{lccc}
        \toprule
        Response & Precision & Recall
        & F1 Score \\ \midrule
        Yes               & 0.79               & 0.94            & 0.86              \\ 
        No                & 0.79               & 0.48            & 0.60              \\ 
        \bottomrule
    \end{tabular}
    \end{minipage}
\end{minipage}
\end{table}

\mysec{Cross Validation with External Datasets}
We also cross-referenced the repositories in \textsc{SciCat} against several existing scientific software datasets from the literature to calculate recall.
Note that because \textsc{SciCat} is not intended to be complete and due to the different sampling criteria and different goals of the various existing datasets, recall is not directly computable. 
For example, Papers with Code contains many replication packages, thus we expect that \textsc{SciCat}'s recall values with respect to it would be low even if \textsc{SciCat} was complete, since our sampling criteria are designed to exclude code dumps as much as possible, which many of the replication packages are. 
Still, we attempt these comparisons to better contextualize our work, by first applying the same filtering criteria we used to identify \textsc{SciCat} candidate repositories before calculating recall values:

\vspace{-1mm}
\begin{itemize}
\setlength\itemsep{3pt}
\item \textit{ Journal of Open Source Software (JOSS):}
JOSS is a peer-reviewed open access scientific journal that focuses on the publication and dissemination of open-source software from any research discipline~\cite{katz2018joss}. 
After scraping 2,200 repositories from the JOSS website~\cite{katz2018joss} and applying our size- and activity-based filters, we were left with 446 unique projects. 
Of these, 314 are part of our \textsc{SciCat} dataset (70\%). 
A closer examination of the missing ones revealed that many (95) were filtered out during the first classification phase (which involved sampling), with the remaining not classified into one of our predefined fields and stack layers during the second classification phase.

\item \textit{National Laboratory Repositories:}
Analysis of open repositories from national laboratory GitHub organizations~\cite{https://doi.org/10.6084/m9.figshare.5328442.v3} resulted in 155 projects meeting our initial filtering criteria (out of 1,910), 
80 of which are in \textsc{SciCat} (47\%).
As with JOSS, most (71) of the 75 missing national lab projects were filtered out during the first classification stage.
Of these 71, further inspection revealed that 59 were categorized as scientific support software or research software only (we did not sample from these; recall Figure~\ref{fig:venn1}), and 12 were not marked with any of our three possible flags or with paper/funding mentions (likewise, we did not sample from these).
Separately, we inspected a random sample of the national-lab repositories that did not meet our initial sampling criteria, and found many examples we would consider scientific software according to our definition, which suggests that our filters could be further relaxed.


\item \textit{Research Software Directory:}
The Research Software Directory~\cite{spaaks2018research} is a content management system for research software containing 411 projects, 184 of which are linked to repositories. 
After applying our filtering criteria, only 19 repositories matched. 
Of these, 14 are in \textsc{SciCat} (74\%), while 4 were filtered out during the first classification step.

\item \textit{Papers with Code:}
From the dataset of 227,820 machine-learning projects by Papers with Code, only 8,547 met our size and activity criteria.
Of these, 4,965 are in \textsc{SciCat} (58\%).
As before, most of the missing ones (3,286 projects) were filtered out during the first classification and sampling stage.
Inspecting these more closely, we discovered that most (about 73\%) were marked as scientific support software or research software without other type labels.
Among the remaining missing ones, we saw many examples of websites, aggregators, courses, and course projects, which do not fit our definition of scientific software.

\item \textit{NSF Soft-Search:}
\citet{brown2023softsearch} identified research projects likely to have produced software while funded by federal grants.
Out of 1,520 entries in the dataset, 88 met our initial criteria, of which 64 were included in our final dataset (73\%).
Among the missing projects, 15 were not labeled as scientific application software during the first classification round and 9 were labeled with layers and fields outside of our scope in the second phase.

\end{itemize}

 

In summary, cross-validation revealed varying degrees of overlap between \textsc{SciCat} and external datasets, with coverage rates ranging from 47\% to 73\%. 
These results demonstrate the diversity of the \textsc{SciCat} dataset, while highlighting the inherent challenges in constructing a comprehensive and fully representative collection of scientific software repositories.
The two main opportunities for future work to increase coverage are relaxing the initial sampling criteria and scaling up the LLM-based processing of README files, whereas we had to resort to sampling.

\vspace{-1mm}
\begin{tcolorbox}[colback=white, colframe=black, boxsep=2pt, left=2pt, right=2pt, top=2pt, bottom=2pt]
In general, \textsc{SciCat} can accurately capture relevant projects within the scientific software domain. In addition, more than 40\% of the projects in our dataset do not mention any publication or funding source in their README (and thus may be hard to identify otherwise).
\end{tcolorbox}
\vspace{-2mm}

\subsection{Limitations}
\label{sec:scicat-limits}

\textsc{SciCat} is incomplete because our sampling criteria for candidate repositories from World of Code (WoC), while derived from established repository mining practices and literature, are relatively strict. 
It also likely misses some scientific repositories hosted individually, since WoC primarily aggregates data from platforms like GitHub and GitLab.
Both limit the generalizability of our findings to a subset of the broader scientific software landscape.
Finally, \textsc{SciCat} is to some extent inaccurate (although our evaluation results above show that accuracy is high) because of the LLM's imperfect ability to label the README files. 
However, all these limitations can be considered \textit{accidental}, i.e., we expect that they can be reduced substantially as LLMs improve, and with more computational resources to process more repositories in and outside WoC. 
More generally, the use of LLMs introduces various known and yet unknown
risks. We, therefore, heavily rely on manual validation of LLM-based classification
via stratified sampling. 

At the same time, a noteworthy \textit{essential} limitation of \textsc{SciCat} is that its construction relies on repository README files.
While, as we show above, this is a good choice when READMEs are present and well written, across public repositories in WoC README files are also often incomplete, vague, or poorly maintained, which can lead to misclassification or the exclusion of important projects. 
This design choice may result in underrepresentation, especially for projects that do not follow standard documentation practices or offer insufficient details about their scientific focus in their README.
Therefore, we expect that a promising direction for future research is to combine a repository content-based approach like ours with an approach based on mining software mentions in research papers~\cite{howison2016software}.


\section{RQ2: Modeling the Longevity of Scientific OSS Projects} 
\label{sec:RQ2}
Next we study the longevity of scientific software repositories in \textsc{SciCat}, understood as the time span from the project's earliest to its latest commit.
Concretely, we label projects as \textbf{\textit{inactive} (abandoned)} if they had \textbf{no commits in the last six months} prior to the end of our data collection,\footnote{We include robustness checks for different operationalizations in our Supplementary Materials. The coefficient estimates vary slightly, as expected, but their direction is consistent.} and record the date of the last commit as the abandonment date.
The commit data are right censored, i.e., projects still active less than six months prior to the end of our observation period may have become abandoned shortly thereafter, or may have remained active to this day (there is more uncertainty about the ``abandonment event'' at the right end of the observation period).
Therefore, we use survival analysis to model the variation in the time to the abandonment event, a standard technique for modeling right-censored time-to-event data~\cite{moore2016applied}.

To compare typical rates of survival across subpopulations, we compute the restricted mean survival time (RMST), which is the mean of the survival time up to some time horizon, i.e., the area under the curve up to a certain point)~\cite{uno2014moving,royston2013restricted}. This is considered a useful summary measure for comparing different survival curves over a given time period. Note that having a time horizon limits the impact of long tails (outliers) on the result; for the purpose of our study, we consider a 15 year window.



\subsection{RQ2a: Variation Across Fields and Layers}

\mysec{Methods}
We start with an exploratory survival analysis of \textsc{SciCat} projects broken down by field and stack layer, in response to RQ2a.
We use the non-parametric Kaplan-Meier (KM) estimator, a robust way to estimate survival functions from time-to-event data.
We will test hypotheses about which factors are associated with variation in these survival probabilities in RQ2b below.


\begin{figure}%
    \centering
    \includegraphics[width=.49\textwidth]
{\detokenize{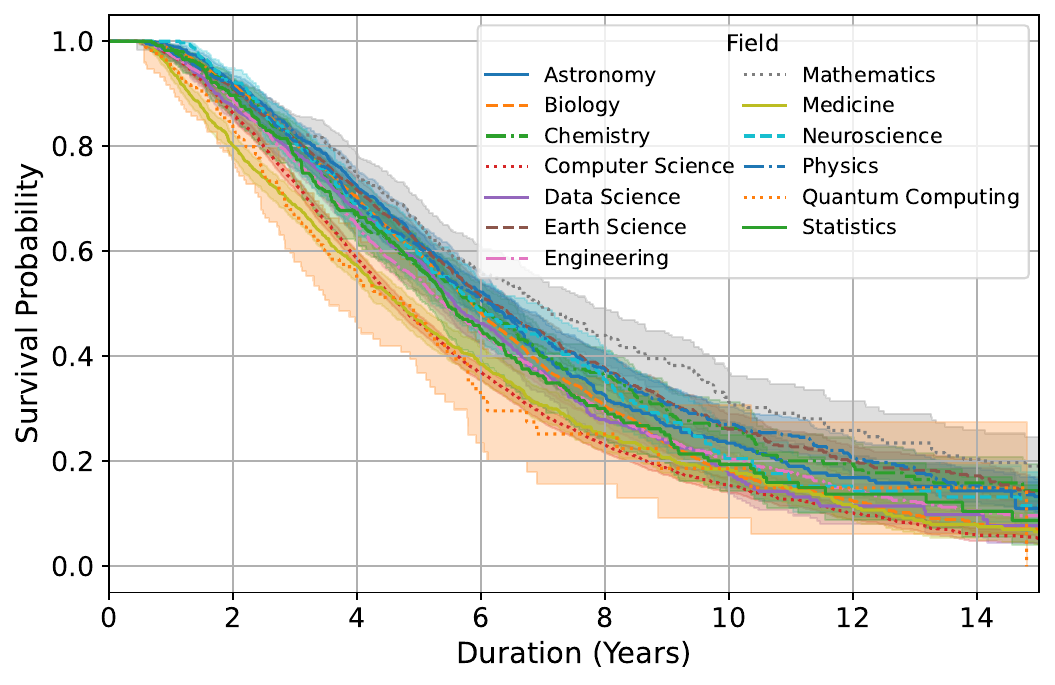}}
    ~
    \includegraphics[width=.49\textwidth]
{\detokenize{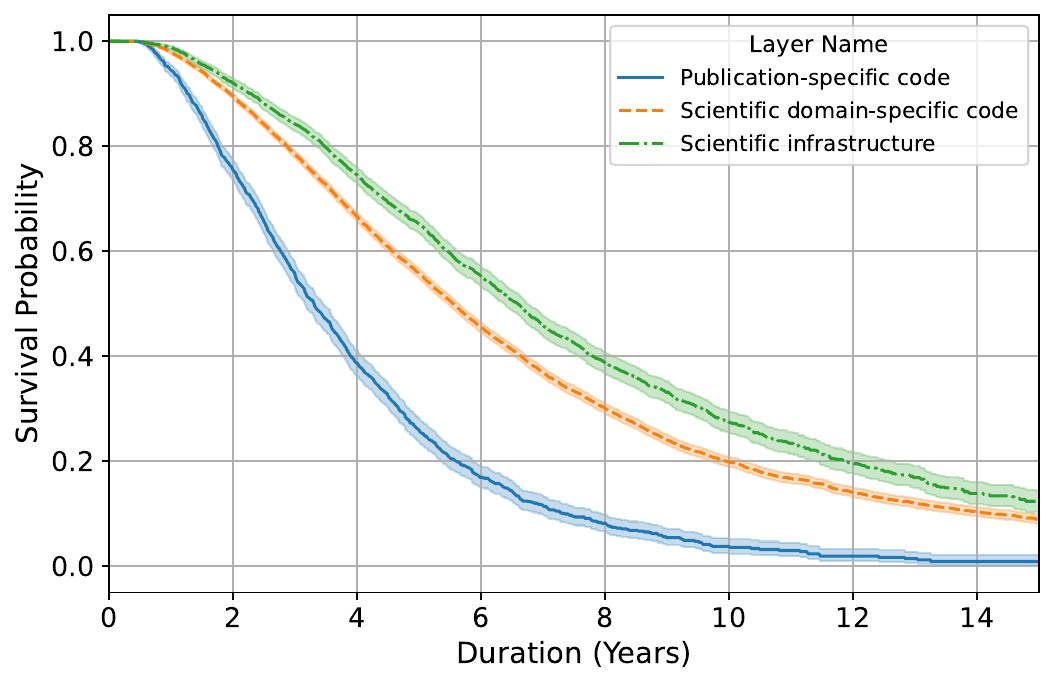}}
    \caption{Kaplan-Meier survival curves by field (left) and layer (right).}%
    \label{fig:survivalplots1}%
\vspace{-3mm}
\end{figure}


\mysec{Results}
The Kaplan-Meier survival plots in Figure~\ref{fig:survivalplots1} offer insights into the longevity of scientific software projects across different domains and layers within the software stack. 
The plot on the right shows survival by layer and compares publication-specific code, domain-specific code, and infrastructure layers. 
The sharp decline in survival probability for publication-specific code in the early years highlights its vulnerability to obsolescence, likely due to its niche use and limited applicability. 
Scientific domain-specific code fares slightly better but still follows a rapid decline, reflecting that while it may have broader usage than publication-specific code, it is still prone to technological shifts or evolving standards. 
In contrast, scientific infrastructure, likely meant to serve more general and long-term needs, does indeed demonstrate a slower, more gradual decline, indicating better longevity. 
This suggests that broad-purpose infrastructure projects tend to persist longer as a result of their foundational role in supporting multiple scientific domains. 

\looseness=-1
The left plot highlights a wide range of survival probabilities depending on the field. 
Fields such as Astronomy, Biology, and Chemistry exhibit relatively steady declines over time, perhaps reflecting their established nature and slower pace of technological change. 
Fields like Computer Science and Quantum Computing experience sharper drops in survival probability, likely due to the rapid advancements and high turnover. 
Medicine and Neuroscience show better longevity, maintaining higher survival probabilities for longer periods, possibly due to the ongoing need for research and development there. 
These survival patterns highlight the variable lifespan of projects in different domains and layers, indicating the complexity of factors that contribute to project longevity.

\vspace{-1mm}
\begin{tcolorbox}[colback=white, colframe=black, boxsep=2pt, left=2pt, right=2pt, top=2pt, bottom=2pt]
Over a 15 year period, the average restricted survival time of scientific OSS projects in our study is 6.44 years. Survival rates in our data varies by field, with mathematics representing the longest average survival (7.91 years) and computer science the shortest (5.74 years). The application layer has significantly shorter survival times than domain-specific (3.94 years vs. 6.52 years), with infrastructure having the greatest average longevity (7.44 years).
\end{tcolorbox}

\vspace{-10pt}
\subsection{RQ2b: Longevity or Survival Analysis:}

\mysec{Methods}
We run a Cox proportional-hazards multiple regression analysis~\cite{RN60} of \textsc{SciCat} projects, which allows us to simultaneously model the relative importance of different factors in explaining the variation in the risk of projects becoming abandoned, according to our six-month inactivity definition.
The semiparametric nature of the Cox model~\citeyearpar{cox1972} also allows us to explore the relative impact of various factors on project longevity without making strict assumptions about the time-to-event distribution. 
This type of regression for survival analysis is standard in numerous domains, including software engineering~\cite{zhou2011does, zhou2012make, zhou2016inflow, valiev2018ecosystem}. 


\begin{table}[t]
\begin{minipage}{\textwidth}
\renewcommand{\arraystretch}{0.5} 
\setlength{\tabcolsep}{4pt} 
\caption{Control (top half) and independent (bottom half) variables measured for our Cox regression.}
\label{tab:vars}
\footnotesize
\begin{tabular}{L{1.75cm}L{3.4cm}L{8cm}}
\toprule
\textbf{Variable}  & \textbf{Description}   & \textbf{Rationale / Hypotheses} \\ 
\midrule
\addlinespace[5pt]
Num.\ Core Authors  & Count of authors with commit counts above the 80th percentile  & While projects may have many \textit{peripheral} contributors, the \textit{core} team, which tends to have the most influence over a project's future, can be very small~\cite{joblin2017classifying, coelho2017modern}. \\ 
\addlinespace[5pt]
Num.\ Commits      & Count of commits & Captures project size and historical activity. Larger projects with more development momentum tend to survive longer~\cite{alves2018understanding}. \\ 
\addlinespace[5pt]
Community Size      & Count of repository forks  &  Similarly to Num.\ Authors, indicates attention to the project and maintenance effort available~\cite{zhou2019fork}. \\ 
\addlinespace[5pt]
Earliest Commit Year & Year of earliest commit & Controls for possible environmental changes in the OSS ecosystem over time~\cite{alves2018understanding}.  \\ 
\addlinespace[5pt]
  Num.\ Defined Packages   & Count of packages defined (e.g., \texttt{package} statements in Java) in the repository 
  & Software intended for reuse is more likely to be packaged for distribution~\cite{decan2019empirical} and intended reusability should increase longevity.  \\
  \addlinespace[5pt]
  Upstream Package Ratio & Fraction of packages used (e.g., as \texttt{import} statements in Java) that are defined in ``upstream'' repositories 
                        & This is a proxy measure for the amount of external upstream dependencies.\footnote{Given the multilingual nature of WoC, these cannot be counted directly.} Projects with more upstream dependencies face more breaking changes~\cite{bogart2021breaking}, package abandonment~\cite{miller2023we}, and other risks. They also tend to accumulate more technical debt, leading to maintenance burdens that increase the risk of abandonment~\cite{dey2019npm, cogo2021deprecation}. \\ 
  \addlinespace[5pt]
  Num.\ Downstream Projects & Count of ``downstream'' projects that import packages defined in the repository & More downstream dependents indicate a larger user base and possible lock-in effects, which may motivate maintainers to keep the project active~\cite{valiev2018ecosystem, dey2019npm}.  \\ 
  \midrule
  \addlinespace[5pt]
  Language           & Dominant language in the repository, based on filename extensions &  Much scientific software is developed in languages like Python and R. Different language ecosystems tend to have different culture and norms~\cite{bogart2021breaking}. \\ 
  \addlinespace[5pt]
  Layer              & Inferred software stack layer (Section~\ref{sec:RQ1}) & Software that plays an infrastructural role for science is expected to be longer-lived~\cite{Hinsen2019}. \\ 
  \addlinespace[5pt]
  Field              & Inferred STEM field (Section~\ref{sec:RQ1}) &  \cite{hasselbring2020open} report significant differences in longevity between research software in different fields, among software with explicit links to publications. \\ 
  \addlinespace[5pt]
  Mentions Paper or Funding  & True if the README was labeled as mentioning papers/funding (Section~\ref{sec:RQ1})  & The combination of financial support and academic recognition should help attract new contributors and funding opportunities \cite{Howison2011, koehler2020better, strasser2022ten}.  \\ 
  \addlinespace[5pt]
  Has Academic Participants & True if any core authors had a .edu-domain email & Captures obvious academic involvement, which can impact software longevity in many ways~\cite{koehler2020better, valiev2018ecosystem}. \\ 
  \addlinespace[5pt]
  Has Government Participants & True if any core authors had a .gov-domain email & Captures obvious government involvement, such as US National Labs. Such involvement may indicate professional (research) software engineers~\cite{koehler2020better, schwartz2024survey}. \\ 
  \bottomrule
\end{tabular}
\end{minipage}
\vspace{-1mm}
\end{table}

In our model we include a number of control variables corresponding to known factors associated with project longevity in the traditional OSS literature, plus several independent variables following our discussion in Section~\ref{subsec:sustainabilityBackground} of factors that may impact the longevity of scientific software differently compared to traditional OSS. 
Table~\ref{tab:vars} lists all the control and independent variables we measured, together with the rationale for their inclusion and our hypotheses, where possible, for their effects. 
All these variables were calculated using WoC data.

The controls include measures of \textbf{software size}, \textbf{project team size}, and \textbf{community size}, as well as measures indicative of the project's position in the global software supply chain based on its \textbf{upstream and downstream dependencies}. 
The independent variables include the three we inferred in Section~\ref{sec:RQ1}, \textbf{field}, \textbf{layer}, and having \textbf{mentions of publications or research funding}, which allows us to formalize the exploratory analysis in RQ1 (field and layer) and contextualize our results in the scarce existing literature (mentions paper or funding).
We also test for differences between \textbf{programming languages}, as their usage tends to differ substantially between scientific fields.
Moreover, we test for effects associated with \textbf{academic and government participation}.
Collaboration between academic and non-academic participants can enrich an open source project, fostering continuous innovation and the transfer of cutting-edge knowledge~\cite{lakhani2003, mockus2002, fitzgerald2006transformation, ghosh2005understanding,colazo2010following}; at the same time, it can be challenging for scientists and engineers to collaborate~\cite{sun2024sustain}, and there may be greater-than-average turnover induced by the academic involvement because of rotating graduate students and postdocs~\cite{valiev2018ecosystem}.
Government participation may come with increased stability and access to resources~\cite{reddy2002}, which could translate to improved software longevity and impact~\cite{weiss2005, sharma2007making, schweik2012internet, ribeiro2020governments}; there are many examples of long-lasting government-funded projects lasting years or even decades~\cite{BOZEMAN2000627, Nelson1993}.

As with all statistical models, we want to establish a relationship between the response and predictors, as statistical models do not demonstrate causal relationships. 
One of the first steps in statistical modeling is to avoid highly correlated predictors, as they increase the errors of the coefficient estimates and often lead to hard-to-interpret results. 
To this end, we computed pairwise rank correlation coefficients between all our variables and eliminated number of authors, where correlation with core authors was above $0.65$.
As multiple comparisons inflate the risk of false discovery, we applied the conservative Bonferroni correction~\cite{sedgwick2012multiple}.

\begin{figure}[t]
   \includegraphics[width=0.9\linewidth, clip=true, trim= 0 40 0 15]{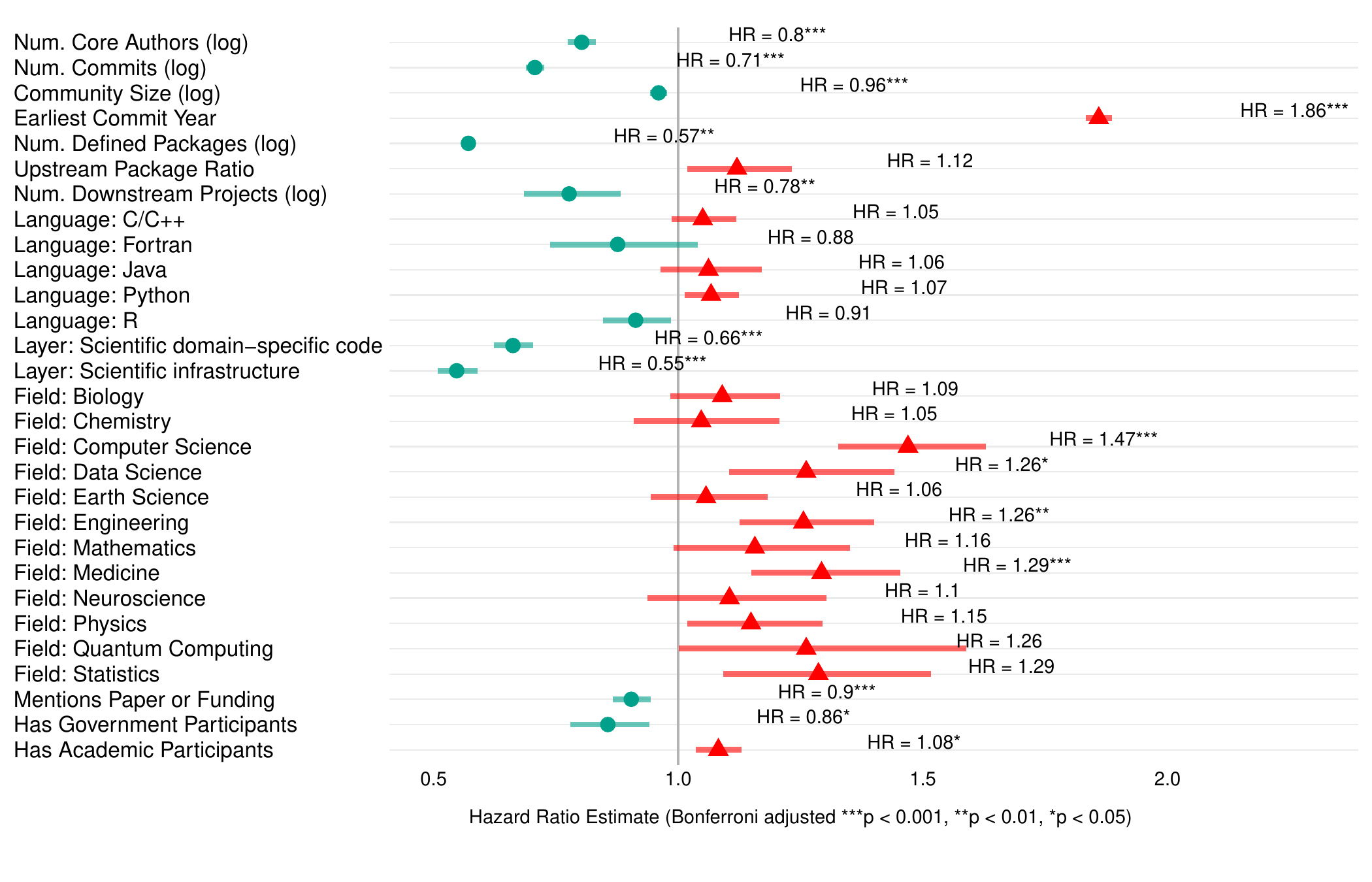}
    \caption{The Cox proportional hazard regression model for scientific software. Out of 18,244 projects, 11,631 had their last commit more than six months before the
   data was collected.\vspace{-0.2cm}}
   \label{fig:ModelScientific}
\end{figure}

\mysec{Results}
Next, we present the findings of our Cox proportional-hazards regression analysis exploring the relationship between the survival times of scientific software repositories in our \textsc{SciCat} dataset and the predictor variables we discussed above.
Figure \ref{fig:ModelScientific} visualizes the estimated \textit{hazard rates} (HR) (exponentiated Cox model coefficients).
It is crucial to note that in the Cox proportional-hazards model, \textit{HR above one indicates an increased likelihood of project cessation, not survival}. 

Looking first at our control variables, we observe effects largely
consistent with prior work.  Larger team size (\textbf{number of
  core authors}), project size (\textbf{number of commits}), and
\textbf{community size} all indicate lower risk of project dormancy
($HR = 0.8, 0.7, 0.96$, respectively; for example, increasing the
log of the core author count by 1 decreases the hazard rate by
20\%), consistent with the understanding that strong and active
community participation enhances OSS projects' longevity.  The
\textbf{number of packages defined} in the project and the
\textbf{number of downstream dependents} also increase longevity
($HR = 0.57$ and $0.78$), indicating that more reusable projects
serving critical roles in the broader ecosystem tend to be
maintained for longer periods.

\looseness=-1
Note that the percentage of projects with at least one
downstream dependent varies with layer in the expected fashion:
11\%, 26\%, and 37\% for publication-specific, domain-specific, and
scientific infrastructure, respectively.  That is, as we transition
from specific publication code to more foundational scientific
infrastructure, projects are more likely to have downstream
dependencies, as expected for infrastructure.  Also note that not
all infrastructure components have downstream dependencies. For
example, runtime environments, workflows, and other tools
categorized as infrastructure may not always provide clear
indications of their use in downstream projects.
It is notable, that the infrastructure projects are longer-lived
even after adjusting for downstream dependents.

While the 95\% confidence interval for the \textbf{upstream project
  ratio} does not overlap with 1, the effect (HR of $1.12$) is no
longer statistically significant after the Bonferroni
correction. If, however, we consider that the correction is very
conservative and the previous consistent findings in the
literature, an increase in \textbf{upstream project ratio} is
likely associated with an increase in abandonment risk. Thus, the
network of upstream dependencies, when large relative to the focal
project's own modules, is associated with shorter project lifespan.

Surprisingly, we note a strong effect associated with the 
\textbf{earliest commit} year: the more recently the
project started, the higher the risk of abandonment ($HR = 1.86$),
indicating perhaps a drastic change in open source culture over the
years. 




Moving on to our main predictors, the model is consistent with the observation from
Section~\ref{sec:RQ1} that scientific infrastructure projects have
the lowest abandonment hazard among the three \textbf{layers} ($HR =
0.55$), followed by domain-specific code ($HR = 0.66$).  
The coefficient for \textbf{scientific domains} represent that
field's contrast with Astronomy (the baseline level, chosen
alphabetically).  
Scientific fields such as Computer Science, Data Science, Engineering, and Medicine
have HR above one, indicating shorter lifespans than comparable projects in Astronomy. 
While Mathematics had the largest marginal survival rate in the
exploratory analysis above in Section~\ref{sec:RQ1}, when adjusting
for other covariates it is trending worse than Astronomy (HR of 1.16
or about 16\% more likely to be abandoned than a comparable
Astronomy project), albeit not statistically significantly after the
Bonferroni correction. 
Our analysis, post adjustment, also did not find statistically
significant differences between Astronomy and Biology, Chemistry,
Earth Science, Neuroscience, Quantum Computing, and Statistics.  
No differences among \textbf{programming languages}
could be observed except for R (which had HR of $0.91$) that was
no longer significant after adjustment.


Projects \textbf{mentioning funding or scientific publications} show longer lifespans ($HR = 0.9$) than comparable projects without such mentions. 
\textbf{Government participation} reduces the hazard rate ($HR = 0.86$), which implies that government involvement brings stability and long-term viability to a project. 
Interestingly, \textbf{academic participation} is linked to a slight increase in hazard ($HR = 1.08$), possibly due to the often short-term or cyclical nature of academic commitments.

\vspace{-1mm}
\begin{tcolorbox}[colback=white, colframe=black, boxsep=2pt, left=2pt, right=2pt, top=2pt, bottom=2pt]
  Like other OSS, scientific software with 
  fewer upstream and more downstream dependencies, larger
  communities, and more core contributors tend to live longer. 
  Government participation and absence of academic
  participation, indications of funding, and
  mentions of scientific publications also tend to increase
  longevity.
  The lower levels of the stack, as expected, survive longer.
\end{tcolorbox}
\section{RQ3: Comparing Longevity of Scientific Software with the Broader OSS Ecosystem} 
\label{sec:RQ3}
For our final analysis we compare the longevity of the scientific software repositories in \textsc{SciCat} to a matched group of non-scientific OSS repositories.


\mysec{Methods}
To estimate if the non-scientific software has different survival
times, we need to conduct a ``natural experiment''. If we 
select a random sample of non-scientific software, we will be
comparing not longevity but precursors of longevity, such as activity.  
The goal for the sample selection process
is, therefore, to be noninformative or ignorable~\cite{LR87} by
which we mean that the resulting sample and population are
compositionally similar on the set of covariates that explain
treatment effect (survival time) variability~\cite{stuart2011use,tipton2013improving}.

For scientific software, we select all projects and we match
the distribution of scientific projects to that of non-scientific
using the following stratified sampling procedure: We
divided our scientific sample based on three variables, namely
number of commits, number of authors, and earliest commit date.
These three give us a gauge for activity, community size, and time,
respectively.  We divided the number of commits into four bins, the
first bin being projects with less than 750 commits, the second bin
with commits between 750 and 1,800, the third bin between 1,800 and
5,000, and the last bin projects with more than 5,000 commits.  We
did the same for number of authors with first bin less than 10
authors, second bin 10 to 25 authors, third bin 25 to 60 and last
bin more than 60 authors.  The logic for choosing these thresholds
was based on our data distribution, so that each bin has roughly
half the count of the previous bin.  That is, the first bin in each
category corresponds roughly to 54\% of the data, the second bin to
27\%, third to 13\% and fourth to 6\%.  Finally, we divided our data
based on their earliest commit date into three bins: projects before
2016, between 2016 and 2018, and the third bin with anything after 2019.
These thresholds were chosen so that the bins have roughly the same
number of projects.  With the explained criteria, we created 48 bins
(4*4*3) and for each bin, we randomly sampled twice the number of
scientific projects from non-scientific projects in OSS using WoC's
MongoDB projects database.  That is, if, for example, there were 500
projects in a bin with 750-1800 commits, 25-60 authors and after
2019, we randomly sampled 1,000 nonscientific projects with these
exact criteria for number of commits, authors, and earliest commit
time.  Having 18,247 scientific projects, we sampled 36,494
non-scientific projects and created a total sample of 54,741
scientific projects and comparable non-scientific projects.

We then estimated a survival model with the specification from RQ2b, 
excluding the domain and layer variables which are not available
for non-scientific repositories, and adding a categorical indicator 
for scientific software to distinguish between the two groups.
\begin{figure}
    \includegraphics[width=0.8\linewidth, clip=true, trim=0 35 0 15]{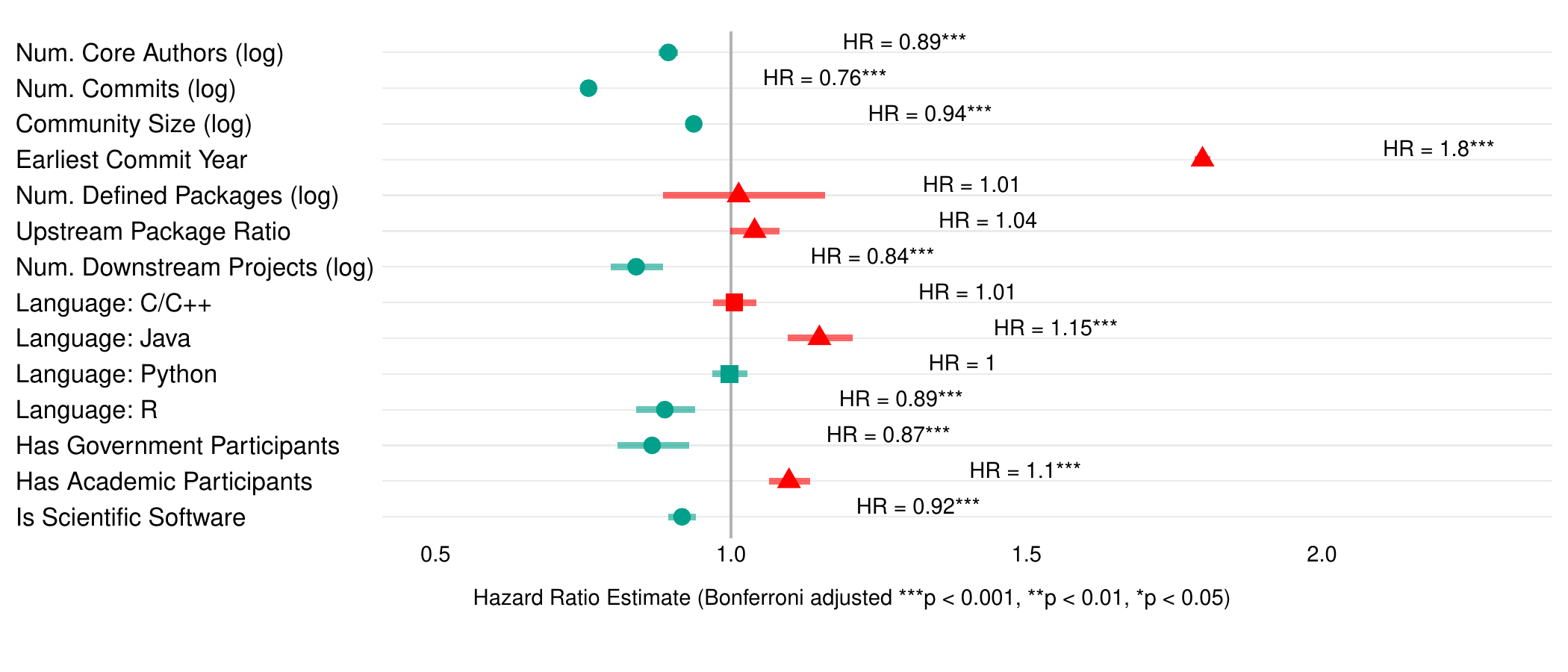}
    \caption{The proportional hazard regression model for scientific and
  matched non-scientific software. Out of 54,710
  projects, 41,069 had last committed more than six months before the
  data was collected.}
    \label{fig:ModelSciAndOSS}
    \vspace{-0.15in}
\end{figure}

\mysec{Results}
Figure~\ref{fig:ModelSciAndOSS} visualizes the estimated hazard
rates for various factors and their statistical significance.
Given the larger sample and a different set of predictors, the
estimated coefficients, as expected, differ slightly from the model used in
RQ2.  Specifically, the model can better discriminate among
\textbf{programming languages} with Java projects having shorter and
R projects having longer lifetimes. Eliminating science-specific
factors from the model makes the number of packages defined in the
project no longer statistically significant. The direction of the
effects stays the same and their interpretation was provided
previously.  The key result of this analysis, though, is given by the
population comparison indicator (\textbf{Is Scientific Software}),
which is has negative $HR = 0.92$ (or about a 8\% lower abandonment
risk), indicating longer survival times.

\vspace{-1.5mm}
\begin{tcolorbox}[colback=white, colframe=black, boxsep=2pt, left=2pt, right=2pt, top=2pt, bottom=2pt]
Scientific OSS projects survive longer than counterparts of the same size
from the same era. 
\end{tcolorbox}

\section{Threats to Validity}
We discussed limitations of our dataset above in Section~\ref{sec:scicat-limits}. 
Here we discuss additional threats to the validity of our analyses in Sections~\ref{sec:RQ2} and~\ref{sec:RQ3}.

\mysec{Construct Validity}
The most notable threat comes from our operationalization of project abandonment which which may overlook ``feature-complete'' projects that no longer need frequent updates but remain valuable and functional.
It is also possible that OSS maintainers return to projects after long breaks~\cite{calefato2022will}, which implies that some of the projects we label as abandoned may resume activity eventually. To assess this threat, we experimented with different thresholds of abandonment, up to 24 consecutive months of inactivity (see Supplementary Materials), with no qualitative change in our findings; that is, breaks are unlikely to invalidate our results.

\mysec{Internal Validity}
First, to ensure the soundness of our regression analysis, we started by modeling a wide range of variables to minimize potential biases from confounding factors, and then reduced that number by excluding highly correlated predictors to improve accuracy of the estimates and to simplify interpretation.
We also corrected data errors, such as times in the future, and checked for outliers or unusual patterns of residuals to minimize undue influence of single observations on the overall model. 
In addition, we used transformations where appropriate to reduce the influence of outliers or to improve interpretability of the results.

Second, since the substantially higher longevity of scientific projects in our sample compared to non-scientific ones is a surprising result, we conducted an additional robustness check, analyzing GitHub star counts as a measure of popularity and attractiveness to contributors~\cite{fang2022damn,borges2018s}; one can expect that more popular projects are less likely to become abandoned. 
Since we did not explicitly match projects on stars (RQ3), there is a risk that our estimated longevity differences between scientific and non-scientific projects can be attributed to differences in star counts.
Therefore, we attempted to control for stars in our regression, but our community size predictor (number of forks) had almost 90\% correlation with star count. 
In conclusion, we exclude this possible alternative explanation for our result.

However, other alternative explanations may still exist as our study is observational.
For example, even though when sampling to select comparable non-scientific projects, projects were matched based on size, activity, and contributors, there may be other predictors of longevity we did not model or match between samples, such as project goals and maintenance practices, the lifecycle of research projects, and changing research priorities. 
More research is needed to investigate these.



\mysec{External Validity}
While this study spans multiple scientific domains, it may not fully represent all areas of research, particularly interdisciplinary fields that use non-standard terminologies. 
Niche scientific communities with distinct documentation practices may also be underrepresented, as would fields outside of the 13 we considered, affecting the generalizability of our conclusions. 
Thus, we advise caution when extending them to other software ecosystems. 
However, this limitation does not diminish the relevance of the findings within the intended scope.

Finally, our study primarily focuses on open-source scientific software with a substantial history of activity, which may overlook important sustainability patterns in less active repositories, such as feature-complete or archival projects. 
We also cannot make claims about proprietary scientific software, which operates under different incentive structures, emphasizing profitability and internal funding, rather than community-driven sustainability. 
As a result, the longevity patterns observed in this study may not fully translate to proprietary or industry-developed software. 
While findings should not be extrapolated to proprietary or archival software without caution, they provide valuable insights into longevity within scientific software ecosystems with community participation.

\vspace{-1mm}
\section{Discussion and Implications} \label{sec:discussion}

Here we highlight implications of our work for research, scientific software, and policymakers.

\mysec{Enhanced Empirical Access To Scientific OSS}
As a major contribution of our study, we showed that a carefully-prompted LLM can identify scientific open-source software based on the contents of the repository README file even when no explicit mentions of research publications or funding are present. 
Our approach can both identify many known examples of scientific software, and also discover many new ones. 
This adds a new tool to a space where better traceability has long been a concern~\cite{howison2016software}, where contemporary efforts to improve traceability exist (using explicit signals such as mentions of the software in publications and of publications in software)~\cite{wattanakriengkrai2022github, istrate2022large}, and where there are still many open questions of interest to practitioners, funders, policy makers, and researchers.
We view our LLM-based approach as complementary. 
We recommend more future work evaluations of when and how an LLM-based approach like ours, which may be less precise but have higher recall than ones based on explicit links, could be combined with the latter to maximize accuracy.

\looseness=-1
There are many open questions about scientific software that our publicly-available \textsc{SciCat} dataset could help answer at scale. 
Researchers could use our dataset to investigate unexplored dimensions of scientific software sustainability, such as the impact of governance models, contributor networks, the evolution of funding sources, and the relationship between sustainability and scientific impact\cite{barabasiGrant}. Meanwhile, we believe our quantiative study lays the groundwork for future qualitative ones that could be based on a theory-relevant sampling of the curated list of projects and/or to further develop the generalizability and applicability of these findings across fields. Of particular interest could be the social sciences and humanities, where software sustainability challenges differ due to funding constraints, project-based development, and the dependence on individual researchers for maintenance~\cite{tucker2022facing}. 

In the process of validating and cross-referencing our collection
with multiple external sources we found a number of drawbacks to
both manually and automatically curated resources. While extracting references to publications can be easily automated,
a large portion of scientific software does not include such
references. Furthermore, a reference to a publication, even if it is
provided, often may not be related to the software itself and may not be an indication that a repository contains scientific
software. Manually curated scientific software collections include
projects that are of educational nature, like tutorials or
documentation. Furthermore, we found most of the projects in such
manually curated collections to be inactive.

\mysec{Supporting Scientific OSS Development}
Our analysis of \textsc{SciCat} projects revealed substantial differences in longevity between scientific OSS at different layers of the software stack, with projects labeled \textit{scientific infrastructure} (i.e., those expected to be better maintained) being indeed the longest lasting. 
It also revealed a positive association between traditional OSS success factors (such as community involvement and having a large and active user base) and longevity. 
We interpret this as good news, in the sense that the same mechanisms that contribute to the success of non-scientific open-source software also seem to impact scientific OSS in predictable ways, despite the different challenges that scientific OSS are subject to.
This adds some evidence to suggest that investing in scientific software can yield long-term benefits for funding agencies, and that programs such as the US National Science Foundation’s Pathways to Enable Open-Source Ecosystems (POSE) are worthwhile, as they support often-overlooked but critical factors such as community participation.

On a practical note, our survival model confirmed that reusability of scientific software is associated with greater longevity -- both the software labeled infrastructure and that with 
many downstream dependent projects tend to be maintained longer. 
This suggests that scientific software developers should aim to design and package their tools in a way that facilitates reuse by other projects, increasing their utility to broader audiences and, in turn, prolonging their lifespan.

The longevity was strongly associated with downstream dependencies,
much more so than with upstream projects. It demonstrates that
carefully measuring usage in downstream projects is important, yet
nontrivial task (in contrast to the identification of upstream
packages). We relied on unique capabilities of the WoC research
infrastructure~\cite{ma2021world} to obtain these measures.

Our analysis revealed two surprising findings.
First, government participation was linked to increased longevity, while academic involvement was linked to reduced longevity, highlighting the need for further research on sustaining scientific software in these different contexts.
Second, projects that mentioned publications or funding in their README tended to last longer.

One possible explanation is visibility: software clearly identifiable as scientific may attract more contributors on platforms like GitHub, where purpose influences engagement~\cite{huang2021leaving}.
Such projects may also be more likely to draw attention from funders and other scientists, increasing their chances of being cited or mentioned in scholarly work.

We recommend future research compare the success of highly visible scientific software (e.g., those in known datasets or frequently cited) with less prominent but equally important infrastructure projects~\cite{nesbitt2024biomedical}.


\mysec{Scientific OSS Is Less Likely To Become Abandoned Than One Might Think}
Our most unexpected finding is that scientific software in \textsc{SciCat} tends to be longer-lived than a matched sample of non-scientific OSS projects. This suggests that while concerns about scientific software abandonment are valid in some cases, they may be overstated when applied broadly.

Although this result holds across different definitions of abandonment and robustness checks (see above), we urge caution in generalizing it beyond our sample, which leans toward larger, collaboratively developed projects rather than smaller or individually developed ones. More research is needed to understand whether scientific software outside \textsc{SciCat} behaves differently.

Still, our findings raise the question of whether scientific software benefits from unique external factors, like academic incentives and funding, that are short-term and unpredictable but largely absent in non-scientific OSS. Ironically, this suggests non-scientific projects might benefit from adopting certain practices common in the scientific realm: tying development to research or business outcomes, promoting use in publications or white papers, and seeking external funding may all help improve sustainability.
\section{Conclusion}\label{sec:conclusion}
\vspace{-1mm}
In our study, we produced and analyzed a large $\sim$18K) and diverse (13 scientific domains) dataset of collaboratively developed OSS scientific software projects as a vehicle for studying scientific software sustainability; we used LLMs with curated prompts to automatically classify project READMEs, checking our results through both manual validation and comparisons with prior datasets and online collections. We analyzed scientific software longevity across this population and compared scientific projects with general open-source ones. We find that scientific software tends to be longer-lived, and factors like fewer upstream and more downstream dependencies, larger communities, government involvement, and the absence of academic participation were linked to increased longevity. 



This curated collection helps to further research
scientific software and results in findings that
can inform science policy related to software
development.  Although our study offers some insight into the
longevity of open scientific software, more work is needed. 
Future enhancements to our collection could enable researchers to assess the role of funding and
institutional support, to include the humanities and social sciences,
and support the study of less active
projects.

\vspace{-1mm}
\section{Data Availability}\label{sec:dataavailability}
\vspace{-1mm}
All data and materials used in this study are publicly available in the replication package.\footnote{\url{https://doi.org/10.5281/zenodo.15278783}} 
\begin{acks}
This work was partially supported by the National Science Foundation NSF Awards 2120429, 1901102, and 2317168. This manuscript has been authored by UT-Battelle, LLC, USA under Contract No. DE-AC05-00OR22725 with the U.S. Department of Energy. The publisher, by accepting the article for publication, acknowledges that the U.S. Government retains a nonexclusive, paid up, irrevocable, worldwide license to publish or reproduce the published form of the manuscript, or allow others to do so, for U.S. Government purposes. The DOE will provide public access to these results in accordance with the DOE Public Access Plan (http://energy.gov/downloads/doe-public-access-plan).
\end{acks}

\clearpage
\bibliographystyle{ACM-Reference-Format}
\bibliography{references}

\end{document}